**Title: eSource for clinical trials: implementation and evaluation of a standards-based approach in a real world trial.**

**Authors:**

Jean-Francois Ethier,[a] Vasa Curcin,[b] Mark M. McGilchrist,[c] Sarah N. Lim Choi Keung,[d] Lei Zhao,[d] Anna Andreasson,[e] Piotr Bródka,[f] Radoslaw Michalski,[f] Theodoros N. Arvanitis,[d] Nikolaos Mastellos,[g] Anita Burgun,[a] Brendan C Delaney.[h]

a. INSERM 1138 Université Paris-Descartes, b. Department of Informatics, King's College London, c. Public Health Sciences, University of Dundee, d. Institute of Digital Healthcare, University of Warwick, e. Division of Family Medicine and Primary Care, Karolinska Institute, f. Department of Computational Intelligence, Wroclaw Institute of Science and Technology, g. Department of Primary Care and Public Health, Imperial College London, h. Institute of Global Health Innovation, Imperial College London.

**Corresponding Author:** Brendan C Delaney

Address: Chair in Medical Informatics and Decision Making, Dept of Surgery and Cancer, Imperial College, St Mary's NHS Trust, Praed St, London, UK, W2 1NY.

e-mail: brendan.delaney@imperial.ac.uk

a) INSERM 1138, eq 22
Université Paris-Descartes
Paris, France
ethierj@gmail.com
anita.burgun@aphp.fr

b) Department of Informatics
King's College London
London, United Kingdom
vasa.curcin@kcl.ac.uk

c) Public Health Sciences
University of Dundee
Dundee, United Kingdom
m.m.mcgilchrist@dundee.ac.uk

d) Institute of Digital Healthcare, WMG
University of Warwick
Coventry, United Kingdom
S.N.Lim-Choi-Keung@warwick.ac.uk
Lei.Zhao@warwick.ac.uk
T.Arvanitis@warwick.ac.uk

e) Division of Family Medicine and Primary Care

Karolinska Institute
Stockholm, Sweden
Anna.Andreasson@ki.se

f) Department of Computational Intelligence
Wroclaw University of Science and Technology
Wroclaw, Poland
piotr.brodka@pwr.edu.pl
radoslaw.michalski@pwr.edu.pl

g) Department of Primary Care and Public Health
Imperial College London
London, United Kingdom
n.mastellos@imperial.ac.uk

h) Institute of Global Health Innovation,
Department of Surgery and Cancer
Imperial College London
London, United Kingdom
Brendan.delaney@imperial.ac.uk

**Keywords:** Electronic Health Records, Clinical Trial, Learning Health System, Interoperability, Operational Data Model

**Word count**: 4786

**ABSTRACT**

**Objective:** The Learning Health System (LHS) requires integration of research into routine practice. 'eSource' or embedding clinical trial functionalities into routine electronic health record (EHR) systems has long been put forward as a solution to the rising costs of research. We aimed to create and validate an eSource solution that would be readily extensible as part of a LHS.

**Materials and Methods:** The EU FP7 TRANSFoRm project's approach is based on dual modeling, using the Clinical Research Information Model (CRIM) and the Clinical Data Integration Model of meaning (CDIM) to bridge the gap between clinical and research data structures, using the CDISC Operational Data Model (ODM) standard. Validation against GCP requirements was conducted in a clinical site, and a cluster randomised evaluation by site nested into a live clinical trial.

**Results:** Using the form definition element of ODM, we linked precisely modelled data queries to data elements, constrained against CDIM concepts, to enable automated patient identification for specific protocols and pre-population of electronic case report forms (e-CRF). Both control and eSource sites recruited better than expected with no significant difference. Completeness of clinical forms was significantly improved by eSource, but Patient Related Outcome Measures (PROMs) were less well completed on smartphones than paper in this population.

**Discussion:** The TRANSFoRm approach provides an ontologically-based approach to eSource in a low-resource, heterogeneous, highly distributed environment, that allows precise prospective mapping of data elements in the EHR.

**Conclusion:** Further studies using this approach to CDISC should optimise the delivery of PROMS, whilst building a sustainable infrastructure for eSource with research networks, trials units and EHR vendors.

**1.1 INTRODUCTION**

The Randomised Clinical Trial remains the standard for approval of new treatments in healthcare. [1] Data standards for research data collection have been formulated by the clinical trials community via The Collaborative Data Standards Interchange Consortium (CDISC) over several decades, with an established pathway for data management from source to submission for regulated clinical trials. Using CDISC standards has led to a steady move away from paper case report forms (CRFs) towards electronic data capture (EDC) systems. Given the rapid expansion of the use of electronic health record (EHR) systems in clinical settings, it has been proposed that EHRs could be the primary point of data entry for a clinical trial. However, direct collection of data into digital form, referred to as eSource, can only be achieved if the EHR is able to support research quality data collection.[2] Good Clinical Practice (GCP) principles need to be adopted to ensure that the requisite standards are in place for eSource, while changes are made to the data collection process and governing regulations to fit in with this electronic context.[3] Moving towards eSource, the Integrating the Healthcare Enterprise (IHE) collaboration (www.ihe.org)[4] has developed a set of profiles including the Retrieve Form for Data Capture (RFD) and Retrieve Process for Execution (RPE), specifying forms and workflow respectively. Several proof-of-concept studies using IHE profiles have been completed.[5]These include STARBRITE, a single site proof of concept implementation within a clinical trial in heart failure patients, that took place nine years ago, without further progress in the field.[6] Complementary to these have been the efforts of the i2b2 community (www.i2b2.org) that has been utilising RedCap software (www.project-redcap.org) to design study forms and collect data directly into i2b2 data warehouses, rather than the EHR.[7]

Within the academic and pharmaceutical trials world there has been a move to 'real world' clinical trials (also known as 'pragmatic' clinical trials) as a means of gathering more representative data, at lower cost, on the likely effectiveness of treatments, to satisfy increased regulatory requirements in this area.[8] Real World clinical trials have simple inclusion and exclusion criteria, are conducted in clinical settings where the treatment will be used, using 'current standard care' rather than placebo as a comparator, and have outcomes collected as a combination of routine clinical contacts and Patient Reported Outcome Measures (PROMs). [9] These outcomes are often used as the basis for an economic analysis. The results of such studies are likely to be closer to the 'in practice' clinical effectiveness and impact of a treatment than the efficacy determined by a typical phase III study against placebo in highly selected subjects. Real world trials are subject to the same effects of co-morbidity, heterogeneity and lack of blinding that accompanies real world use of a treatment. As real world trials take place in routine clinical settings, a compelling case can be made for eSource, using the existing EHR to support the study.[10] It has been proposed that embedding research into routine EHR systems, could automate a substantial part of the trial's screening process.[11] Eligibility criteria can be partially tested against EHR patient data and electronic case report forms (eCRFs) can be pre-filled with data present in the EHR, in order to minimise unnecessary manual entry. In addition, clinical data collected within a trial should be made available in the EHR, in order to enhance routine clinical care and safety monitoring.[12][13]

Real world trials have not yet progressed to using eSource by default, still requiring a large investment in data collection and validation.[14] Closing this step would go a long way to providing an end-to-end 'research and learning' continuum for a Learning Health System (LHS), where research and knowledge translation are routinely transacted via ICT systems. Use of robust data standards, such as the CDISC suite, when interacting with EHRs, are essential to the operation of the LHS in order to overcome the 'silo of excellence' culture prominent in healthcare research, and lower the barrier to entry for traditional clinical environments.[15] In this paper, we describe an approach to embedding clinical trial functionality within the EHR systems, enabling the pre-population of eCRFs directly from the EHRs, and the potential of recording of CRF data, collected during research, within the EHR system. This paper describes the methods we adopted to create semantically enriched and model-based extensions to existing standards, how we implemented these in live EHR systems, and how we validated and then evaluated the technical functionality of the approach in a live clinical trial as part of a large European research programme.

## 1.2 The TRANSFoRm infrastructure

The European FP7 TRANSFoRm project aimed at developing an infrastructure for a Learning Health System in European Primary Care (www.transformproject.eu),[16] a major workstream of which was directed at developing eSource connectivity for randomised controlled trials (RCT). Primary Care represents the ultimate low-resource, heterogeneous, highly distributed environment, especially when the multi-language, multi-health system dimensions of Europe are added.

Under the IHE approach, only single EHR systems have been used to deploy standard forms, pre-populated with limited EHR data. In each case, the study data collection requires the use, or at least conformance, to a minimum set of data elements defined in CDASH (Clinical Data Acquisition Standards Harmonization),[17] and the EHR system needs to be capable of collecting and managing the CRFs, possibly with custom extensions. This approach requires a large academic centre for conducting trials, with support for a complex IT infrastructure, and the close participation of EHR vendors. TRANSFoRm, on the other hand, had to consider the requirements of multi-site, multi-system data collection with a low resource overhead, such that the LHS can encompass a range of healthcare system at various levels of IT maturity, not just large academic centres. Therefore, what is required is a readily extensible framework, enabling researchers to define clinical data elements to research standards like CDISC and to semantically align them to native EHR data. TRANSFoRm has taken an approach of using existing CDISC standards, but referencing a core data model, expressed as an ontology, to provide a flexible and more streamlined approach. The requirements established for TRANSFoRm are shown in Table 1.

| **Clinical study requirements:** |
| --- |

1. Prevalent and incident case identification from live EHR systems in primary care

2. Real time alerting when a case is identified via EHR

3. Pre-population of CRFs displayed within the EHR user interface

4. Data capture in EHR fulfilling the eSource requirements of GCP (noting that if blinded, treatment allocation must remain concealed). These include data provenance and validation of data capture and transfer accuracy.

5. Patient Related Outcome Measure data captured electronically and stored in the EHR for Safety monitoring

6. Data provenance – towards compliance with 21 CFR Part 11 and European regulation

7. Full evaluation in 5 EU member states and in the context of a real world RCT

| **Technical requirements:** |
| --- |

8. To use ontologies to maintain models of meaning for the LHS, terms being bound to clinical concepts

9. To use CDISC foundational standards including the Operational Data Model and (ODM), and the Study Data Model (SDM)

10. To enable connection to multiple country, multiple language, multiple vendor systems with minimal vendor input

| **Vendor requirements:** |
| --- |

11. Standard Terminology used in EHR

12. Sample EHR data set available for testing

13. Represent local database metadata as a model (DSM) and map to TRANSFoRm Clinical Data Integration Model (ontology)

14. Availability of an Application Programming Interface (API) and a demo installation for testing (or the vendor builds the DNC functionality into their system)

Table 1: Requirements for embedding RCTs in an EHR as part of a LHS.

## 2. MATERIAL AND METHODS

### 2.1 General approach

TRANSFoRm took a unified approach to workflow and data integration for the entire project, described previously.[16] The requirements and workflow of the research process are first expressed using the Clinical Research Information Model (CRIM),[18] a domain-specific implementation of CDISC's Biomedical Research Integrated Domain Group (BRIDG).[19] It is, for example, the role of CRIM to direct when in the workflow a query should be used for retrieving patient study data from aggregated EHR data repositories. When required, in order to extract data from clinical sources, the unified interoperability framework separates the stable domain information from the heterogeneous data sources to achieve structural and semantic interoperability between different actors in the LHS (clinical investigators, EHRs, researchers, CRFs).[20–22] It works by binding structural and terminological models (of both the domain and the sources) in order to derive the full semantic meaning of clinical data.[21,23] The clinical primary care domain is specified, using the Clinical Data Integration Model (CDIM),[24] an ontology that enables users to work with data and express queries using neutral clinical concepts, without needing any knowledge about the specific schema of the target data source. CDIM offers a unified view of the primary care domain and CDIM has been developed as a realist ontology, all of whose classes have instances in the real world, and works in conjunction with medical terminologies that provide concrete instantiations. CRIM and CDIM models together specify the data flow through TRANSFoRm's infrastructure.

### 2.2 ODM and eSource

The CDISC Operational Data Model (ODM) is a vendor neutral, platform-independent XML format for interchange and archive of clinical study data, designed to facilitate regulatory-compliant acquisition, archive and interchange of data and metadata for clinical research studies (http://www.cdisc.org/odm). ODM's `<FormDef>` element captures eCRF composition and structure, with `<ItemGroupDef>` elements used to group related items (e.g. a systolic blood pressure measurement value, the unit of measure and the time at which it was measured), with `<ItemDef>` elements containing specific item metadata. As shown in Figure 1, we used `<ItemGroupDef>` to reference research data queries expressed using CRIM, and `<ItemDef>` to define clinical data elements referencing the CDIM ontology. As both these elements are contained within `<FormDef>`, the necessary model constraints are applied. Our key requirements were pre-population of forms from existing EHR data and controlled data elements in the EHR. In order to embed an EHR data extraction request into ODM, `<ItemGroupDef>` was extended by adding a `<QueryId>` child element, containing a unique identifier linking the item group with the corresponding query. CDIM is used to annotate `<ItemDef>` through its `<Alias>` element, which allows binding to an external model using the context attribute (e.g., CDIM_2.2) and the value attribute (e.g., CDIM_000070). As an ontology-based mediation, the pre-population query does not contain source specific structural information and is used for every source. Once the embedded Data Extraction Queries have been translated for the specific source by the TRANSFoRm interoperability framework and executed, the results are annotated with CDIM concepts and placed in the proper `<ItemDef>`, as identified by the `<Alias>` element. In this way, TRANSFoRm's dual-level modelling enables data interoperability between EHR patient data and the ODM. PROMs are collected using a separate smartphone and web application using the TRANSFoRm ODM extension to specify the mobile and web data collection.[25]

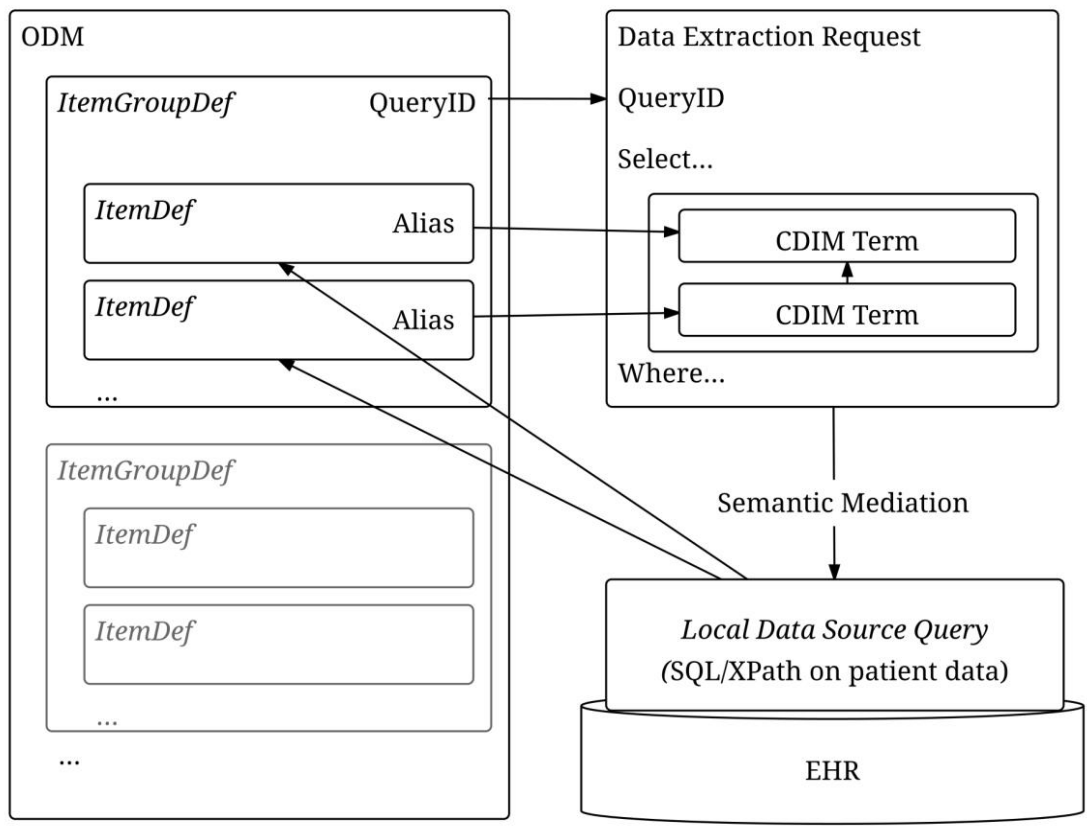

Figure 1: CDISC and TRANSFoRm eSource interactions

## 2.3 Validation

A clinical use case was developed to guide the development, validation and evaluation of TRANSFoRm. The gastro-oesophageal reflux disease (GORD) study aimed to assess the effectiveness of TRANSFoRm in patient recruitment and answer the following clinical question: "*What gives most symptom relief and improvement in quality of life (QoL) in patients with GORD, on demand or continuous use of proton pump inhibitors?*".[26] Preparations took place between late 2013 and early 2015 in Poland, beginning with the integration of the vendor system (mMedica from Asseco Poland S.A.) and the TRANSFoRm platform through a single platform component, the data node connector (DNC), which brokers communication with the TRANSFoRm study system and other platform components. The intention of the validation study was to ascertain the accurate functioning of the TRANSFoRm tools and to carry out a Good Clinical Practice (GCP) certification. GCP is a requirement for the use of data collection systems in regulated clinical research, in the US known as 21 CFR 11.[27]

A simulated study with 10 process scenarios was designed.[28] The software was installed in practices and data collection scenarios carried out by Polish clinicians and TRANSFoRm staff acting as patients. The training plans were also designed and tested with pilot users. The installation and regular operation of TRANSFoRm components, including data collection tools, TRANSFoRm study system and the data node connector, were documented through a set of Installation Qualification, Operation Qualification and Performance Qualification tests, all performed on the pilot trial site. The development teams involved in software production were themselves assessed in terms of training, software quality assurance procedures and institutional policies.

## 2.4 Evaluation

It has been suggested that a robust way of evaluating methodological innovation in clinical trials is the 'SWAT', or 'Studies within a Trial' design, where a second randomisation allocates study subjects or sites to a alternative methods of delivering the main trial. Following this approach.[29] We aimed to conduct a mixed-methods evaluation of the TRANSFoRm eSource method as a nested cluster randomised trial embedded fully within an RCT (700 subjects, individually randomised). The studies were conducted across five countries and five different EHR systems (EudraCT trial number 2014-001314-25), according to a published protoco.[26] Ethical approvals were obtained in the UK, Netherlands, Belgium, Poland and Greece. The Study Sponsor was the Karolinska Institute, Sweden. The aim of the evaluation study was to compare the TRANSFoRm tools to standard methods for opportunistic clinical trial recruitment in primary care, which are largely based on searches of patient records conducted within a given EHR system, without any 'real time' alerts.[30] The Clinical RCT results will be published separately. The primary outcome of the study was recruitment rate, based on an effect size of the TRANSFoRm system increasing recruitment of subjects by 75% (RR 1.75, based on a baseline of 20% in the control arm. We aimed to measure the denominator of eligible subjects across centres via searches of the EHR at study end, as a very high proportion of primary care encounters are coded, enabling efficient measurement of patients meeting the eligibility criteria in each arm across centres.[31][32] The secondary outcomes were recruitment per week per site and data completeness.

The practices in the TRANSFoRm arm had the TRANSFoRm software installed in the practice and were trained in using the software by TRANSFoRm staff. The eCRF system used in the control arm consisted of a basic web-based electronic case report form with built in randomization algorithm identical to that used in the TRANSFoRm eCRF tool, but with no EHR interaction, thus lacking the capabilities to support patient recruitment by flagging eligible patients, and prepopulating forms with EHR data. Patients completed the PROM on paper and the questionnaire was sent to the local study coordinator for data entry. (Table 2) Differences in recruitment rate between the TRANSFoRm arm and the control arm were calculated using Wilcoxon matched-pairs signed-ranks test and the pairing from the randomization of the practices to the TRANSFoRm vs the control arm was used. Differences in completion rate were analysed using two sample test of proportions.

| | TRANSFoRm | Control |
|---|---|---|
| *Recruitment (consecutive)* | Flagging patient in EHR | Manual identification of eligible patients from the EHR |
| *Informed consent* | Signed paper | Signed paper |
| *Randomisation* | Automatic in eCRF | Automatic in eCRF |
| *Data Elements* | Standardised via CDISC ODM and CDIM | No standardisation |
| *CROM collection* | eCRF integrated with EHR and prepopulated with EHR data | Web based eCRF with no pre-population |
| *PROM collection* | Web/Smart Phone application also based on an ODM xml document | Printed questionnaires distributed by practices, prepaid envelopes provided. Data entered manually into database |
| *Monitoring* | Reporting workbench | Manual |
| *Data saved* | Recoverable from EHR | Not saved to EHR |
| *Provenance* | Traced via system | Not traced |

Table 2: Conditions for TRANSFoRm system and control sites (CROM= Clinical Related Outcome Measures, PROM = Patient Related Outcome Measures)

## 3. RESULTS

### 3.1 Implementation of the research platform components and workflows with vendors

We worked with five EHR vendors, practices using the EHR system were then approached and asked to participate. Eight to ten practices per vendor were included in the study, and altogether 36 practices participated in the study: 8 in Greece (University of Crete TRANSHis variation), 10 in Poland (Asseco Mmedica), 8 in the Netherlands (Dutch TRANShis) and 10 in UK (In Practice Systems Vision3). In the UK 10 practices using The Phoenix Partnership's SystmOne could not take part as the integration with TRANSFoRm was not completed on time. The system required on-site configuration via third-party practice ICT support that was hard to engage. The Belgian vendor's EHR system deployment was delayed for commercial reasons and there were no practices using the system to recruit. Vendor's involvement consisted of providing an XML data source model describing their patient data schema, as well as providing some form of API for communication with the DNC – the latter we found already present in all the EHRs we recruited. The effort involved in this work, over and above that already being carried out by the EHR vendor, consisted of some administration of non disclosure agreements, answering queries around the use of the API and conformance testing of the DNC prior to its installation in sites.

The TRANSFoRm Study System (TSS) is centrally hosted at a secure location and holds the study information and protocols, defined via ODM files. It also acts as the research repository for the collected eCRF data. The coordination of study activities at the local level is performed by the data node connector (DNC) components, which sit locally to the EHR instances. All research data capture operations (e.g., eligibility checking), eCRF pre-population and eCRF completion, are orchestrated and performed by the DNC. Thus the data flow between the EHR and the DNC remains local to the EHR and only the data identified for research purposes is sent to the research repository in line with the project's security and data protection framework.[33] The DNC can pull data from the TSS but the TSS cannot push data to nor pull data from the DNC as initiator of the communication. The latter is important as often the DNC will sit behind an organisational firewall in a clinical setting. The DNC is started with the host EHR system and obtains from the TSS information about the currently active study protocols and their eligibility criteria. The generic study definitions are then translated by the Semantic Mediator (SM) component into locally executable queries. When a patient arrives for a consultation, their record is sent to the DNC as an XML document, where it is checked for eligibility. The TRANSFoRm platform does not mandate a specific structure or model for that file, which usually corresponds to the EHR system's native format.

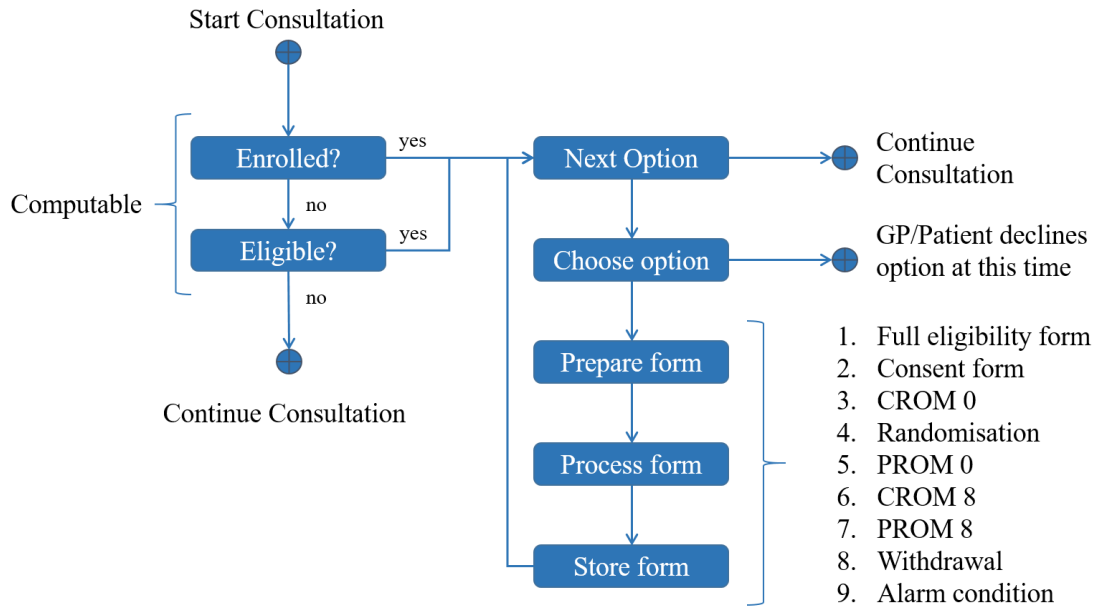

Figure 2: TRANSFoRm study workflow for the GORD trial.

TRANSFoRm workflow is shown in Figure 2. When a presenting patient is found to be potentially eligible for a study, the DNC notifies the clinician of the eligibility via a pop-up message requesting completion of eligibility checks and consent/randomisation. Thereafter when the recruited patient presents at the practice, the DNC retrieves the appropriate eCRF forms from the study system, transported as HTML forms parameterised for pre-loading and storage of field values, together with the corresponding CDISC ODM document container with the ClinicalData section parameterised to store the data values entered into HTML fields. The generation and parameterisation of the HTML and ODM documents is performed by the TRANSFoRm Study System based on a pre-established ODM to interface translation, OID, QueryID and CDIM alias. Using this information, the DNC can correctly pre-load form fields by applying the queries to the patient data extract and inserting the resulting values at the corresponding place in the form. The pre-loaded HTML form can then be presented to the clinician for validation and entry of data items that have not been pre-loaded. The form can either be embedded into the EHR or accessed through a web browser.

Once approved, the form is submitted to the DNC where data is inserted in the ODM file. The DNC then sends the ODM document containing the responses from the form to the TSS for research activities. It also sends the associated ODM files and HTLM forms to the EHR for auditing purposes and reviewing by clinicians at a later date, if required. The EHRs currently partnering with TRANSFoRm do not store the form field data as individually coded facts, but as a single artefact which can be viewed as a whole. While the mappings between the patient data extract and CDIM could be used to support granular transfers back to the EHR, the vendors preferred not to explore this aspect in the first version. Data elements that were pre-populated and their availability in each system are shown in Table 3. Data flow and document examples are shown in supplementary material on-line.

| Data element | Ontology ID | Asseco_nMedica (PL) | InPS_Vision (UK) | TransHIS (NL) TransHIS (GR) |
|---|---|---|---|---|
| Patient clinical research ID symbol | CDIM/3 | Y | Y | Y |
| Gender role | OMRSE/7 | Y | Y | Y |
| Human birth instant | CDIM/7 | Y | Y | Y |
| Health encounter instant | CDIM/79 | N | Y | Y |
| Physician practice† | OMRSE/17 | [1] | [1] | Y |
| Diagnostic conclusion | OGMS/73 | **Y/ICD10** | **Y/ReadV2** | **Y/ICPC** |
| Diagnostic conclusion instant | CDIM/12 | Y | Y | Y |
| Symptom | OGMS/20 | N | Y | N |
| Mass measurement datum | CDIM/68 | Y | Y | Y |
| Mass measurement instant | CDIM/67 | Y | Y | Y |
| Mass measurement unit label | CDIM/100 | [kg] | [kg] | Y |
| height measurement datum | CDIM /71 | Y | Y | Y |
| height measurement instant | CDIM /70 | Y | Y | Y |
| height measurement unit label | CDIM /88 | [cm] | [cm] | Y |
| sys BP measurement datum | CDIM /73 | Y | Y | Y |
| sys BP measurement instant | CDIM /102 | Y | Y | Y |
| sys BP measurement unit label | CDIM /84 | [mmHg] | [mmHg] | Y |
| dia BP measurement datum | CDIM /74 | Y | Y | Y |
| dia BP measurement instant | CDIM /101 | Y | Y | Y |
| dia BP measurement unit label | CDIM /83 | [mmHg] | [mmHg] | Y |
| formulated pharmaceutical item | CDIM /37 | **Y/ATC** | **Y/MultiLex** | **Y/ATC** |
| Rx instant | CDIM /105 | Y | Y | Y |
| Laboratory test | OGMS/56 | N | **Y/ReadV2** | **Y/LOINC** |
| Laboratory measurement scalar value | CDIM/32 | N | Y | Y |
| Laboratory confirmation instant | CDIM/29 | N | Y | Y |
| Laboratory measurement unit label* | CDIM/81 | N | N | Y |

Table 3: Pre-populated data elements and coverage by system

**Notes:**

[] indicates implicit values for a data element which are directly specified in the structural mapping model.

* Failure to identify an explicit laboratory unit label had no consequences since TRANSFoRm-1 did not perform unit conversion.

† Physician practice was always accessed as 'current physician'

## 3.2 Validation study

In order to achieve GCP certification, the TRANSFoRm system underwent a series of tests, including Installation Qualification, Operational Qualification and Performance Qualification. These tests established the functional correctness of the system, with respect to the requirements and specification, and also establishing the integrity of the data that is output from the system. The non-functional aspects examined included training materials, technical support, skill level of the development teams, and software quality assurance procedures. Post-installation support was provided by members of the TRANSFoRm and vendor teams covering the use of the updated vendor software and supporting TRANSFoRm software. Most issues arose from the sequencing of forms to be filled within the EHR system. In the cases where this was implemented by the vendor, it is necessary to check that forms will be submitted in the correct order. Extensive logging by the TRANSFoRm DNC meant that a full record of these issues could be maintained and there was less reliance on GP reporting to understand these issues. A full description of the validation related to usability is published elsewhere.[28]

## 3.3 Evaluation study

The number of recruited patients in the TRANSFoRm and control arm in all four localities and in total is presented in Table 4. The total number of recruited patients exceeded 600 and was very similar between the TRANSFoRm and the control arm. Greece and Poland stood for the vast majority (96%) of all recruitments.

|  | TRANSFoRm arm | Control arm | Total |
|---|---|---|---|
| *Greece* | 122 | 121 | *245* |
| *Netherlands* | 10 | 6 | *16* |
| *Poland* | 156 | 177 | *333* |
| *UK* | 5 | 3 | *8* |
| **Total** | **293** | **307** | **600** |

Table 4: Recruitment of subjects by site and arm.

Greece and Poland were used to compare recruitment rates between the TRANSFoRm arm and the control arm. Eight pairs of practices were available for analysis. The total number of patients recruited was very similar between the arms. The total number of eligible patients was higher in the TRANSFoRm arm, as there was one practice that had a very long recruitment time and consequently a high number of eligible patients according to the EHR. The study was powered to detect an increase in the recruitment rate from 20% to 35%. The average recruitment rate was 43% in the TRANSFoRm arm and 53% in the control arm with a large range in recruitment rate between practices. There was no significant difference in recruitment rate between the TRANSFoRm arm and the control arm nor was there a statistically significant difference in number of recruited patients per week between the two arms (mean TRANSFoRm=2.84 recruited patients per week vs mean control 2.39 recruited patients per week, p=0.67).

All localities were included in the comparison of completion rate in the TRANSFoRm arm compared to the control arm. In the TRANSFoRm arm, 85% of those with a first Clinical Reported Outcome Measure (CROM) had a filled 2nd CROM, while in the control arm this was true for 71 % (p<.001). In the TRANSFoRm arm, 61% of patients with a first PROM had also filled out 2nd PROM, as compared to 100% in the control arm (p<.001). Hence, the TRANSFoRm tool supported CRF data collection significantly better while the manual distribution of questionnaires was superior to the use of the mobile/web app for PROM data collection.

## 4. DISCUSSION

TRANSFoRm has shown that the process of integrating clinical trial process and data management into the EHR can be based on CDISC standards without demanding significant workload from the EHR vendors. A recent review of embedding RCTs for effectiveness in EHRs, that referenced some of our earlier work that informed TRANSFoRm,[13][10] concluded that 'substantial re-engineering of the EHR is required to allow for trial workflow.[34] Our work shows how barriers to adoption can be lowered and increasing the uptake of the LHS, whilst retaining compatibility through use of standards. The key components of this approach are not specific to TRANSFoRm, but extensions to the CDISC approach as follows:

1. An ontology or a data model appropriate to the domain to bind to data elements, better defining their model of use in the clinical domain (CDIM)

2. Reference to ontology of EHR data mappings, mostly these will be references to tables in the EHR, but a detailed clinical data element definition and underlying model can be used if those definitions are in clinical use

3. A study system for managing artifacts and models, definitions etc and transacting workflow (TSS)

4. A local Data Node Connector for linking the TSS to EHR systems.

None of these components are specific to a particular platform and can be replaced by alternative versions (e.g., locally developed) that conform to the  TRANSFoRm models.

The adoption of CDIM to represent a shared 'model of meaning' allows the separation of definition from implementation.[21] ODM is maintained as the key standard, with references binding item definitions to CDIM and embedding the research meaning as a precisely defined query, structured and guided by CRIM. CDIM, in conjunction with relevant terminologies, allows expression of precise, complete and fine-grained clinical concepts as required by the LHS - answering the needs of users as well as catering to data sources with varying granularity. This simple method is anchored into a foundation model (CDIM) that allows for a high level of granularity and precision while supporting various logical operators as well as covering a wide range of terminologies, including the UMLS. The outlined approach is generalizable to other domains as different domain specific models can be created. We re-used higher level concepts to ensure future compatibility by using BRIDG for CRIM and building CDIM from middle level ontologies.[35,36] A number of approaches have been taken to allow the creation of various data elements 'defined sets' with different level of granularity and detail (ISO11179 Ed3,[37,38] CDISC CDASH,[17] ISO13606,[39] CIMI [5]). While this flexibility is essential to define data elements purely for research purposes, when applied to clinical interoperability, this can quickly lead to a profusion of overlapping data elements with minuscule variations. This is especially problematic when put in context of organizing data flows between the research data structures and the EHRs. Mappings need to be created between EHRs and data elements. All these slight variations can confuse and complicate mapping creation and maintenance. Moreover, to work effectively, links between the EHRs and the data elements need to be established prospectively in order to avoid the need to create a new mapping each time a new data element is derived with a slightly modified definition. Our approach circumvents this problem by using ODM, defining data elements via CDIM, and mapping them into the EHR's native terminology, passing the resulting terms to ODM.

As described earlier, an alternative approach to linking to EHR systems has been taken by IHE and the CDISC Healthcare Link Initiative (HCL) whereby the process of integration is undertaken by EHR vendors.[4] The TRANSFoRm approach should not be seen as a competitor to IHE, but an attempt, in the context of an academically-led project to streamline the process of using CDISC standards for small vendors such as those found in primary care and specialist clinical areas.  Table 5 compares the TRANSFoRm and HCL/IHE approaches. The two approaches are not mutually exclusive. Both the TRANSFoRm and IHE approaches can co-exist by using the `<ItemDef>` element as a pivot by using two aliases: one to CDASH and one to CDIM. Maintaining clinical meaning via an ontology should be seen as desirable in the context of the LHS, where a basic reasoning capability is important in maintaining coherence across a distributed research and translational system. [40,41]

The recent completion of SHARE by CDISC, creating a single repository of both CDISC standards and artefacts from forms to data elements, offers potential to develop an integrated solution whereby CDIM and TRANSFoRm data elements and queries could be made available via SHARE. This would require careful consideration of CDISC subscriptions and the need to cover a variety of industry, academic and clinical users.

| Requirement | HCL | TRANSFoRm |
|---|---|---|
| *Form specification* | ODM | ODM |
| *Research CDE definition* | CDASH | External model (e.g. CDASH) with mapping to CDIM Ontology[35,36] for pre-populated elements. |
| *Research CDE storage and distribution* | CDISC SHARE (ISO11179) | Not implemented could be using CDISC SHARE |
| *Research CDE mapping to clinical DE* | Data Element Exchange (DEX) | CDIM ontology referenced by ItemDef Alias |
| *Pre-population specification of query* | SDM (xpath) via DEX | CDISC Study Data Model (xpath) via pre-specified queries referenced by ODM ItemGroupDef QueryID |
| *Pre-population extraction of EHR data* | Retrieve Process for Execution and HL7 Clinical Research Document | Via Data Node Connector and EHR API |
| *Semantic mapping* | CDASH – restricted code set | TRANSFoRm Terminology Service (LexEVS) augmented by manual term selection and binding |
| *Display of CRFs* | Retrieve Form for Data Capture Profile – proforma implemented by EHR system | Via Data Node Connector and EHR API |
| *Data storage from CRFs* | Archive | TSS (vie Data Node Connector) |
| *Audit and change control* | Archive | Open Provenance Model, provenance trace recorded with operation of tools.[42,43] |
| *Security and authorisation* | Within EHR | TSS (inherited from local authorisation) |

Table 5: Key differences between HCL and TRANSFoRm.

(HCL - CDISC Healthcare Link; ODM – Operational Data Model; CDE – Common Data Element; CDASH – Clinical Data Acquisition Standards Harmonization; CDISC SHARE – Shared Health and Clinical Research Electronic Library; MDR – Metadata Repository; DEX – IHE Data Element Exchange; SDM – Study Design Model; RPE – IHE Retrieve Protocol for Execution; CRD – IHE Clinical Research Document; RFD – IHE Retrieve Form for Data Capture; TSS – TRANSFoRm Study System)

In the RCT we recruited more than 600 patients bringing us close to the goal of 700 recruited patients. However, the meta-RCT evaluation study became underpowered compared with our initial plans as no data was contributed to the analysis from the UK (2 sites, 20 practices) or The Netherlands and 8 rather than the 20 planned pairs of practices included in the analyses.

The TRANSFoRm approach was successfully validated and shown to conform with the necessary GCP requirements to conduct the evaluation study. A very strong reactivity meant that all the Polish and Greek sites recruited more efficiently in both arms than we had expected. We had decided to compare recruitment rates with active sites in the TRANSFoRm and control arms rather than include the effect of start up delays in the eSource sites, as we viewed this as a one off set-up rather than an issue with future studies. The trials-within-trials approach to evaluation of methodological innovations in clinical trials is likely to face problems with power unless a basket of RCTs are used across each innovation, the clustering effects of RCT protocols being taken into account in subsequent analysis.

A defining principle of a Learning Health System is that it is universal, encompassing both new state-of-the-art research environments and the traditional clinical settings with no advanced informatics infrastructure. Our goal therefore is to be able both to define meaning at the system level and also enable incorporation of legacy systems where there is not the resource to develop and maintain separate infrastructures for research within the clinical system. Development of an open and transparent approach to using 'data transfer' standards such as ODM and FHIR, along with ontologies and mappings between detailed clinical terminologies or models, coupled with robust provenance will form the basis for the consistent clinically-rich data standards necessary for the operation of the LHS at scale. It is only through such inclusive approaches, that we shall fulfil on the promise of the LHS and achieve its wider take-up.

## AUTHORSHIP STATEMENT

JFE, AB, MM, BD and VC participated in the development of CDIM, the unified framework and its applications. MM and LZ, with the help of SLCK, JFE and TNA, directed the development of the infrastructure for the implementation of the DNC. AA and NM developed the use case and participated in the validation study. PB and MR developed the TSS and the tools for the patients. JFE, BD, VC and SLCK helped to draft the manuscript. All authors critically reviewed the manuscript and approved the final version. BD was Scientific Director of TRANSFoRm and guarantees the manuscript.

## ACKNOWLEDGMENTS

We would like to thank Rebecca Kush, Michael Ibara and Sam Hume from CDISC for their insightful comments on drafts of this paper.

## COMPETING INTERESTS

None

## FUNDING

This work was supported in part by the European Commission—DG INFSO (FP7 247787) and partially supported by the European Commission under the 7th Framework Programme, Coordination and Support Action, Grant Agreement Number 316097, ENGINE - European research centre of Network intelliGence for INnovation Enhancement (http://engine.pwr.edu.pl/). The RENOIR project - Reverse EngiNeering of sOcial Information pRocessing - leading to this publication has received funding from the European Union's Horizon 2020 research and innovation programme under the Marie Skłodowska-Curie grant agreement No. 691152.

and AN for VM of RA. Search for FDA Guidance Documents - Part 11, Electronic Records; Electronic Signatures — Scope and Application.

2014;**53**. doi:10.3414/ME13-02-0024

**FIGURE LIST**

Figure 1: CDISC and TRANSFoRm eSource interactions

Figure 2: TRANSFoRm study workflow for the GORD trial